\documentclass[aps,pra,twocolumn,showpacs,preprintnumbers,amssymb,superscriptaddress]{revtex4}
\usepackage{graphicx}
\usepackage{dcolumn}
\usepackage{bm}
\usepackage{amsmath}
\usepackage{amsfonts}
\usepackage{verbatim}
\newcommand{\ket}[1]{|{#1}\rangle}
\newcommand{\bra}[1]{\langle{#1}|}
\newcommand{\oket}[1]{\overline{|{#1}\rangle}}
\newcommand{\obra}[1]{\overline{\langle{#1}|}}
\newcommand{\ketbra}[2]{|{#1}\rangle\langle{#2}|}
\newcommand{\braket}[2]{\langle{#1}|{#2}\rangle}
\newcommand{\obraket}[2]{\overline{\langle{#1}|}{#2}\rangle}
\newcommand{\braoket}[2]{\langle{#1}\overline{|{#2}\rangle}}
\newcommand{\cL}{{\mathcal L}}
\newcommand{\cK}{{\mathcal K}}
\newcommand{\cP}{{\mathcal P}}

\newcommand{\Tr}{{\rm Tr}}
\newcommand{\abs}[1]{\left|{#1}\right|}
\newcommand{\av}[1]{\left\langle #1 \right\rangle}
\newcommand{\br}[1]{\langle #1|}
\newcommand{\ke}[1]{|#1\rangle}

\newcommand{\kb}[2]{\ke{#1}\br{#2}}

\newcommand{\pt}[1]{\left( #1 \right)}
\newcommand{\pq}[1]{\left[ #1 \right]}
\newcommand{\pg}[1]{\left\{ #1 \right\}}

\begin{document}


\title{Quantum jumps induced by the center-of-mass motion of a trapped atom}

\author{J. M. Torres}
\affiliation{Instituto de Ciencias F\'isicas, Universidad Nacional Aut\'onoma de
M\'exico, Cuernavaca, Morelos, Mexico
}

\author{M. Bienert}
\affiliation{Abteilung f\"ur Theoretische Quantenmechanik, Universit\"at des Saarlandes, Campus E 2 6, 66041 Saarbr\"ucken, Germany}

\author{S. Zippilli}
\affiliation{Departament de F\'{\i}sica, Universitat Aut\`onoma de
Barcelona, E 08193 Bellaterra, Spain}
\affiliation{Fachbereich Physik and research center OPTIMAS, Technische Universit\"at Kaiserslautern, D-67663 Kaiserslautern, Germany}

\author{Giovanna Morigi}
\affiliation{Departament de F\'{\i}sica, Universitat Aut\`onoma de
Barcelona, E 08193 Bellaterra, Spain}
\affiliation{Abteilung f\"ur Theoretische Quantenmechanik, Universit\"at des Saarlandes,Campus E 2 6, 66041 Saarbr\"ucken, Germany}

\begin{abstract}
We theoretically study the occurrence of quantum jumps in the resonance fluorescence of a trapped atom. Here, the atom is laser cooled in a configuration of level such that the occurrence of a quantum jump is associated to a change of the vibrational center-of-mass motion by one phonon. The statistics of the occurrence of the dark fluorescence period is studied as a function of the physical parameters and the corresponding features in the spectrum of resonance fluorescence are identified. We discuss the information which can be extracted on the atomic motion from the observation of a quantum jump in the considered setup. \end{abstract} \date{\today}

\maketitle

\section{Introduction}

Quantum jumps are one of the earliest concepts in quantum mechanics, which were introduced to denote the atomic transitions between electron bound states due to photon absorption or emission as predicted by Bohr~\cite{Bohr}. This terminology is now used to designate the observation of a sudden change in the fluorescence signal of laser-driven atoms, which is associated with the occurrence of transitions between electronic states~\cite{qjumpsexp1,qjumpsexp2,qjumpsexp3}.
Since their first observation~\cite{qjumpsexp1,qjumpsexp2,qjumpsexp3}, quantum jumps have been extensively discussed in the literature. For an overview we refer to the reviews \cite{Plenio98,CarBooks,Leibfried}. 
Quantum jumps have inspired theoretical descriptions and methods in quantum optics, such as quantum trajectories~\cite{Cohen86,Carmichael91,Molmer,Plenio98}, and are at the basis of several theoretical proposals for post-selected quantum state preparation, some examples are found in Ref.~\cite{StochasticCooling,Almut,Cirac97}, and for detecting the quantum state of trapped atoms~\cite{Vogel,Cirac97}. 
Most recent experiments used quantum jumps in order to demonstrate single photon absorption by single trapped atoms~\cite{Schuck10}, and analyzed their statistics in order to demonstrate quantum correlations between the absorbed photon and a second, entangled photon revealed at a detector~\cite{Piro10}. Further experimental work used quantum jumps in the signal monitoring the number of photons inside a resonator as the basic element for implementing feedback loops, restoring the quantum state of a cavity mode~\cite{Dotsenko09}. These milestone experiments demonstrate once more that quantum jumps can be a powerful tool for monitoring and manipulating the quantum state of physical systems. 

Quantum jumps have been extensively discussed in connection to the preparation and measuring of the quantum state of a trapped atom~\cite{StochasticCooling,StocCool:exp,Vogel,Leibfried}. In a suitable setup their analysis can provide information on the dynamics and the state of the atomic center of mass degree of freedom. Information on the quantum state of the atoms can be also extracted by the spectrum of resonance fluorescence, which, when taken at the steady state of laser cooling, reveals the relevant scattering processes characterizing the dynamical steady state~\cite{Bienert06,BienertTorres07} and can provide a measurement of the cooling rate and steady state phonon number occupation~\cite{Lindberg86,Cirac93,Bienert04,Raab00,Phillips97}. 

For the detection of quantum jumps one usually monitors the fluorescence light of a strongly driven atomic transition. When due to a additional weak coupling the electron jumps to a third, metastable state, the sudden stop of the fluorescence reveals the occurrence of the jump.
The properties in the resulting spectrum of resonance fluorescence have been theoretically analyzed in Ref.~\cite{Plenio95} and measured in~\cite{Tamm2000}. General features in the emission and absorption spectrum of three-level systems have been theoretically investigated in Refs.~\cite{Narducci1990,Manka1991}, and experimentally confirmed~\cite{Gauthier1991,Stalgies1996}. 
During the last decades, three-level systems have become a paradigmatic system for observing quantum interference~\cite{Arimondo:1996,Harris:1997,Fleischhauer:2005}. Signatures thereof can manifest themselves in the spectrum of resonance fluorescence~\cite{Aspect1984, Manka1991, Zhou:1997, Martinez:1997}.

\begin{figure}
\begin{flushleft}
{\bf a)}\\
\end{flushleft}
\vspace{-1cm}
\includegraphics[width=3.5cm]{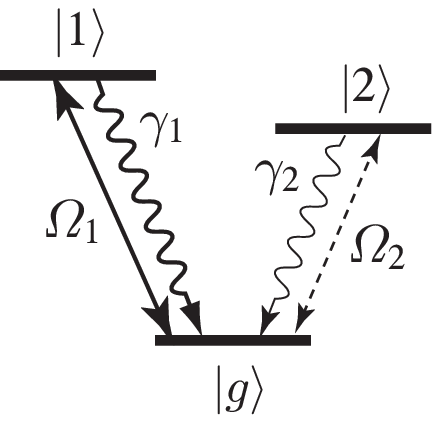}
\begin{flushleft}
{\bf b)}\\
\end{flushleft}
\vspace{-.7cm}
\includegraphics[width=7.5cm]{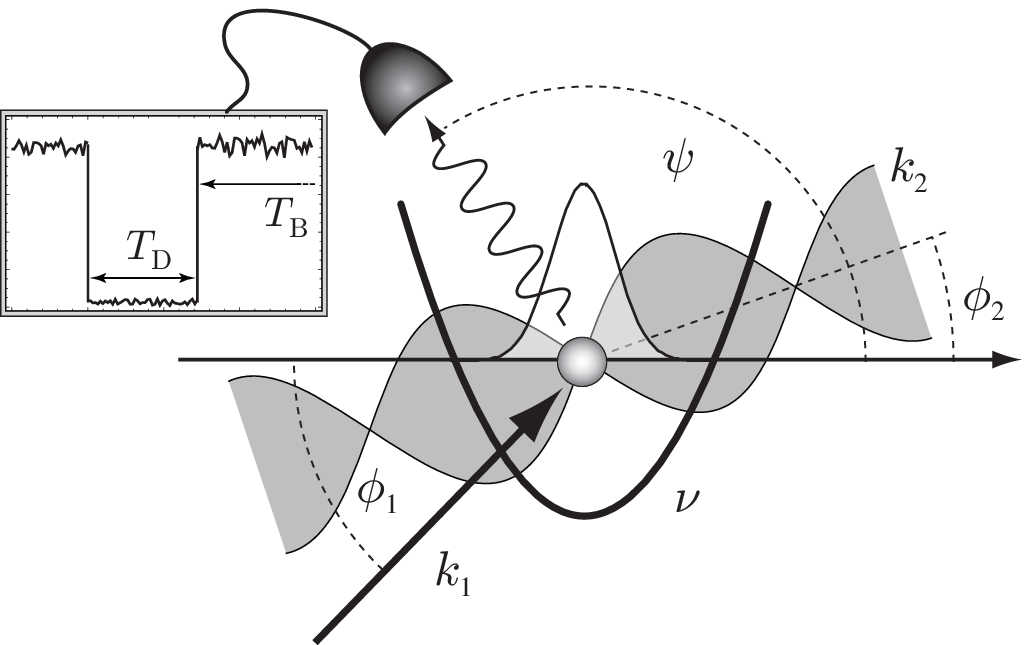}
\caption{
\label{fig:model}
A single atom is confined in a harmonic trap and laser-cooled. a) Relevant levels, showing the dipole transitions $\ket g \leftrightarrow \ket 1$ and $\ket g \leftrightarrow \ket 2$ which are driven by lasers with Rabi frequency $\Omega_1$ and $\Omega_2$, respectively. The state $\ket g$ is stable, and excited states $\ket 1, \ket 2$ decay into it with rates $\gamma_1$ and $\gamma_2$. In this paper $\gamma_2\ll \gamma_1$. b) Setup: The harmonic potential with frequency $\nu$, confines the atomic center of mass motion which is assumed to be along the $x$-direction. The laser coupling the transition  $\ket g \leftrightarrow \ket 1$ is a running wave at wave vector $k_1$ forming the angle $\phi_1$ with the $x$-axis, as illustrated by the black arrow in the figure. The laser coupling the transition  $\ket g \leftrightarrow \ket 2$ is a standing wave at wave vector $k_2$ forming the angle $\phi_2$ with the $x$- axis. In the figure it is represented by the shaded region with sinusoidal envelope. The trap center coincides with one node of the standing wave. A detector collects the photons emitted at an angle $\psi$ with the axis. Inset: Temporal behavior of the resonance fluorescence, consisting of bright and dark periods at time scales $T_{\rm B}$ and
$T_{\rm D}$, respectively. A quantum jump to a dark period corresponds to a transition to state $\ket 2$, and is due to atomic fluctuations around the trap center at a node of the intensity of the laser coupling to state $\ket 2$.}
\end{figure}

In this work we address the question what are the spectral properties associated with quantum jumps, carrying information on the center-of-mass motion of a trapped atom. In particular, we establish the connection between the time dependent fluorescence and the corresponding spectrum originating from the quantum motion of the atomic center of mass. For this purpose, we analyze a setup, constituted by a single trapped atom driven by laser fields which cool its center-of-mass motion, such that the occurrence of quantum jumps in the atom's fluorescence is associated with a change of its center-of-mass motion by a single vibrational excitation. The configuration of electronic levels and the setup are sketched in Fig.~\ref{fig:model}. The telegraphic noise, detected in the fluorescence, is analyzed both in time and in frequency, thereby providing an extensive description of the quantum jump process induced by the quantum fluctuations of the atom's center-of-mass motion. We identify and characterize the spectral features associated with the quantum jumps, which reveal a change of the atomic motional state by one vibrational excitation. Our analysis shows that the spectral components connected with the motion-induced quantum jumps are proportional to the ratio between the dark and bright time scales of the intermittent fluorescence, and scale with the temperature of the laser-cooled atom.

The article is organized as follows. In Sec.~\ref{Sec:2} the setup and the corresponding theoretical model are introduced. In Sec.~\ref{Sec:3} the time scales of bright and dark periods in the resonance fluorescence are evaluated using the delay function~\cite{Cohen86,Plenio95}, and the basic steps leading to the evaluation of the spectrum of resonance fluorescence, are summarized. Moreover, the results are discussed in this section. In Sec.~\ref{Sec:5} the conclusions and a discussion on possible outlooks are presented. The appendices report the details of the calculations at the basis of the results.

\section{The System}
\label{Sec:2}

We consider an atom of mass $M$ whose center-of-mass motion is confined by a harmonic trap. The atom is driven by lasers, and the relevant electronic levels, which are involved in the dynamics, are the stable state $\ket g$ and two excited states $\ket 1$, $\ket 2$ at frequencies $\omega_1$ and $\omega_2$, respectively. The level scheme is sketched in Fig.~\ref{fig:model} a). States $\ket 1$ and $\ket 2$ are excited by means of lasers and decay radiatively into the ground state $\ket g$ with rate $\gamma_1$ and $\gamma_2$, respectively, where we assume that $\gamma_2\ll\gamma_1$. This configuration has been considered for observing quantum jumps in atomic systems: a long-living excitation of state $\ket 2$ leads to interruption of the fluorescence along the transition $\ket g\leftrightarrow\ket 1$~\cite{Itano90,Plenio95}.
In the present setup, however, there is no excitation along the transition $\ket g \leftrightarrow \ket 2$ if the atom is a point like particle, since the laser field driving this transition is a standing wave and the trap center coincides with one of its node, see Fig.~\ref{fig:model} b). Hence, only the mechanical effects of light will induce dark periods in the resonance fluorescence signal.

We now discuss the mechanical effects of the light on the atomic motion. In the considered scheme, the laser driving the transition $\ket g\leftrightarrow\ket 1$ is a running wave with frequency $\omega_{{\rm L}1}$ and wavevector $k_1$, which Doppler cools the atomic motion~\cite{Eschner03}. The laser driving the transition $\ket g\leftrightarrow\ket 2$, instead, is a standing wave with frequency $\omega_{{\rm L}2}$ wave vector $k_2$, where one of its nodes coincides with the center of the trap, as shown in Fig.~\ref{fig:model} b). Excitations of the state $\ket 2$ hence occur due to fluctuations of the atom's position about the center of the trap. In this work we consider that the atomic motion is in the Lamb-Dicke regime, such that the extension of the center-of-mass wave packet is much smaller than the laser wave length and the mechanical effects are hence a small correction to the motion. In this regime, interruptions in the fluorescence on the transition $\ket g\leftrightarrow\ket 1$, quantum jumps, will then occur due to a mechanical excitation or de-excitation of the atomic center of mass motion.

\subsection{Theoretical model}

The dynamics of the density operator $\varrho$, describing the internal and external degrees of freedom of the atom, is governed
by the master equation
\begin{equation}
\frac{\partial\varrho}{\partial t} = \cL\varrho,
\end{equation}
where the Liouville operator $\cL$ is defined as
\begin{equation}
\cL\varrho=\frac{1}{i\hbar}\left[H,\varrho\right] + \cK\varrho,
\label{eq:fullmastereq}
\end{equation}
with the Hamiltonian $H$, giving the coherent evolution of the system, and the relaxation superoperator $\cK$ describing spontaneous emission.
In detail, the Hamiltonian can be decomposed into the sum of the terms
\begin{equation}
H = H_{\rm ext}+ H_{\rm int} +V,
\label{eq:hamiltanH}
\end{equation}
consisting of the center-of-mass motion in the trap, which we assume here to be along the $x-$axis,
\begin{equation}
H_{\rm ext}=\hbar\nu(a^\dagger a+1/2)\,,
\end{equation}
with the annihilation and creation operators $a$ and $a^\dagger$
of a vibrational excitation at energy $\hbar\nu$. Furthermore, it contains the electronic states' energy,
\begin{equation}
H_{\rm int}=\hbar\delta_1\ketbra 11+\hbar\delta_2\ketbra 22\,,
\end{equation}
which are reported in the frame rotating
at the frequencies of the lasers, such that $\delta_j=\omega_j-\omega_{{\rm L}j}$. The laser-atom interaction in electric dipole approximation has the form
\begin{alignat}{1}
V= &\hbar\frac{\Omega_1}{2}\left[\ketbra 1g e^{i k_1 x\cos\phi_1 }
+{\rm H.c.}\right]\nonumber\\
+&\hbar\frac{\Omega_2}{2}\sin\left(k_2 x\cos\phi_2\right)\left[\ketbra 2g
+{\rm H.c.}\right],
\label{eq:interactionW}
\end{alignat}
with the position $x=\xi(a+a^\dagger)$ of the atomic center of mass motion, and
$\xi=\sqrt\frac{\hbar}{2 M \nu}$. The interaction includes the mechanical effects
associated with the absorption and emission of a photon of the laser, which are weighted by the projection $k_j\cos\phi_j$ of the lasers' wave vectors on the $x$-axis.
Finally, the superoperator describing spontaneous decay takes the form
\begin{equation}
\cK\varrho = \sum_{j=1,2}\frac{\gamma_j}{2}\left(2\ketbra gj\tilde\varrho_j\ketbra jg-\varrho\ketbra jj-\ketbra jj\varrho\right),
\label{eq:Kterm}
\end{equation}
where
\begin{equation}
\tilde\varrho_j = \int\limits_{-1}^{1} du\,
w(u)  e^{-i k_j x u}\varrho e^{i k_j x u}.
\label{eq:rhotilde}
\end{equation}
The weight function $w(u)$ is the normalized dipole radiation pattern, giving the probability of emitting a photon at an angle $u=\cos\theta$ with the motional axis. In the following we will take $w(u)=3(1+u^2)/8$, corresponding to the transitions $\ket g\leftrightarrow \ket 1,\ket 2$ being characterized by the selection rule for the angular momentum $\Delta J_z=\pm 1$. Furthermore, we will assume the Lamb-Dicke regime, corresponding to taking the Lamb-Dicke parameter $\eta_j = \xi k_j$ such that it fulfills the relation~\cite{Stenholm86}
\begin{equation}
\eta_j = \sqrt{\frac{\hbar k_j^2}{2 M \nu}}\ll 1,
\label{eq:smalleta}
\end{equation}
together with the sufficient condition $\eta_j\sqrt{\langle n\rangle}\ll 1$, where $\langle n\rangle=\langle a^\dagger a\rangle$ is the mean number of phonons. The mean number of phonons $\langle n\rangle$ is reached by an interplay of the mechanical effects of both lasers. The laser driving the transition $\ket g\leftrightarrow \ket 1$ is tuned to the red side of the atomic resonance, $\delta_1=\gamma_1/2$, and performs Doppler cooling on the atom. In absence of the second laser, the photons scattered along this transition, contain the information about the steady state of Doppler cooling. We also assume that the second laser realizes sideband cooling, such that absorption of a photon along the transition $\ket g\to \ket 2$ corresponds with largest probability to a de-excitation of the atomic motion by one phonon~\cite{Eschner03,Stenholm86,Cirac92}. Although it is not essential for our study, we have optimized the parameters in order to minimize the final temperature of the combined laser cooling. A brief discussion of the cooling dynamics and of the choice of parameters for optimizing the steady-state temperature is reported in App.~\ref{app:cooling}. Below we introduce the expansion of the Liouvillian $\cL$ in powers of the Lamb-Dicke parameters.

\subsection{Lamb-Dicke expansion of the Liouvillian and spectral decomposition}

Master equation~\eqref{eq:fullmastereq} can be formally solved using the spectral decomposition of the Liouvillian $\cL$, namely, the left- and right eigenelements at eigenvalue $\lambda$ obeying
\begin{alignat}{1}
\cL\hat\varrho^\lambda = \lambda\hat\varrho^\lambda,
\quad \check\varrho^\lambda\cL = \lambda\check\varrho^\lambda.
\end{alignat}
Orthonormality and completeness are defined with respect to a scalar product given by the trace, such that ${\rm Tr}\{\check\varrho^\lambda\hat\varrho^{\lambda'}\}=\delta_{\lambda,\lambda'}$. Correspondingly, the projectors $\cP^\lambda$ can be introduced, whose action on an operator $X$ is
\begin{equation}
\cP^\lambda X=\hat\varrho^\lambda \Tr\{\check \varrho^\lambda X\}
\end{equation}
such that the completeness relation can be expressed as
\begin{equation}
\sum_\lambda P^\lambda = 1.
\label{eq:completp}
\end{equation}
We note that the steady state is a right eigenelement of the Liouvillian $\cL$ at eigenvector $\lambda=0$, while the corresponding left eigenelement is  $\check\varrho^{\lambda=0}=1$~\cite{Englert2002}.

Eigenvalues and eigenvectors can be evaluated using the Lamb-Dicke expansion. The Lamb-Dicke expansion of Liouvillian $\cL$, Eq.~\eqref{eq:fullmastereq}, up to second order gives
$\cL=\cL_0+\cL_1+\cL_2+\dots$, where the subscripts label the order in the expansion and the individual Liouvillian terms take the form~\cite{Cirac93}
\begin{alignat}{1}
  \cL_0 \varrho &= \cL_{\rm E}\varrho + \cL_{\rm I}\varrho\nonumber\\
&=\frac{1}{i\hbar}[H_{\rm ext},\varrho] +
\left(\frac{1}{i\hbar}[H_{\rm int} +
V_0,\varrho]+\cK_0\varrho\right)\label{eq:Li},\\
\cL_1\varrho&=\frac{1}{i\hbar}[V_1 x,\varrho]\label{eq:Lfirst},\\
\cL_2\varrho&=\frac{1}{2i\hbar}[V_2 x^2, \varrho] + \cK_2\varrho,
\end{alignat}
for the zero, first and second order terms, where we have used
\begin{equation}
V_n=\left.\frac{\partial^n V(x)}{\partial x^n}\right|_{x=0},
\end{equation}
and
\begin{alignat*}{1}
\cK_0\varrho&=\sum_{j=1,2}
\frac{\gamma_j}{2}\left(2\ketbra gj\varrho\ketbra jg-\varrho\ketbra jj-
\ketbra jj\varrho\right),\\
\cK_2\varrho&={\mathcal W}_2\sum_{j=1,2}
\frac{\gamma_j k_j^2}{2}\ketbra gj (2 x\varrho x-
x^2\varrho-\varrho x^2)\ketbra jg,
\end{alignat*}
where ${\mathcal W}_2=\int\limits_{-1}^{1} {\rm d}u \, w(u) u^2$.

In zero order, the internal and external degree of freedoms are decoupled, as one can also see in Eq.~(\ref{eq:Li}): Liouvillian $\cL_{\rm E}$ describes the equation of motion of a harmonic oscillator,  while Liouvillian $\cL_{\rm I}$ gives the internal dynamics of an atom at rest. Correspondingly, in zero-order perturbation theory the eigenvalues read
\begin{equation}
\lambda_0 = \lambda_{\rm E} + \lambda_{\rm I},
\end{equation}
where $\lambda_{\rm E}=i\ell\nu$ ($\ell = 0,\pm1,\pm2,\dots$) are the eigenvalues of $\cL_{\rm E}$ and $\lambda_{\rm I}$ are the (nine) eigenvalues of $\cL_{\rm I}$. The steady state, right eigenvector at $\lambda_0=0$, takes the form
\begin{equation}
\varrho_0^{\rm st} = \rho_{\rm st}\mu_{\rm st}
\label{eq:stst}
\end{equation}
with the density matrix $\rho_{\rm st}$ defined for the internal degrees of freedom and obeying $\cL_{\rm I}\rho_{\rm st}=0$, while $\mu_{\rm st}$ is the density matrix for the center-of-mass degrees of freedom, which solves the effective equation for laser cooling~\cite{Cirac93,Bienert04}. Projectors on the corresponding subspaces,
\begin{equation}
\cP_0^\lambda = \cP_{\rm E}^{\lambda_{\rm E}}\cP_{\rm I}^{\lambda_{\rm I}},
\end{equation}
are a product of externals projectors
\begin{equation}\label{eq:projectors}
\cP_{\rm E}^{\lambda_{\rm E}=i\ell\nu} X =
\sum_n \ketbra nn X \ketbra{n+\ell}{n+\ell}
\end{equation}
and internal projectors $\cP_{\rm I}^{\lambda_{\rm I}}$ of $\cL_{\rm I}$. The internal projectors $\cP_{\rm I}$ are constructed by the left- and right eigenelements to $\lambda_{\rm I}$, $\check\rho_{\rm I}^{\lambda_{\rm I}}$ and $\hat\rho_{\rm I}^{\lambda_{\rm I}}$, of $\cL_{\rm I}$, see App.~\ref{app:esintern}. Note that the external subspaces are infinitely degenerated. Higher order projectors $\cP_1$, $\cP_2$,\dots and eigenvalues $\lambda_1$, $\lambda_2$,\dots can be calculated using standard techniques of perturbation theory and can be found in \cite{Bienert04,Bienert06}. It turns out that $\lambda_1=0$ and $\cP_0^\lambda\cL_1\cP_0^\lambda=0$ since $\cL_1$ does not couple states of subspaces with same eigenvalue $\lambda_0$~\cite{Bienert04,Bienert06}.

\section{Quantum Jumps and spectrum of resonance fluorescence}
\label{Sec:3}

In this section we identify the characteristic time scales, determining the duration of the bright and dark periods in the resonance fluorescence as a function of time, and then determine the spectrum of intensity of the emitted light.

\subsection{Quantum jumps: Dark and bright periods}
\label{sec:qj}

We now focus on the fluorescence signal as a function of time. We show that, studying light scattering by the atoms at steady state of laser cooling, the fluorescence exhibits bright periods, due to light scattering along the transition $\ket g \to \ket 1$, and dark periods which are due to the shelving of the electron in the state $\ket{2}$. The transitions to the state $\ket 2$ can only take place due to the finite extension of the motional wave packet of the atom at the node of the laser standing wave. The corresponding processes are therefore of higher order in the Lamb-Dicke expansion and associated with a change of the atom's center-of-mass state. In the Lamb-Dicke regime, the time scales characterizing the bright and dark periods of the fluorescence are well separated: Denoting by $T_B$ and $T_D$ the time scales of bright and dark periods, respectively, then $T_B\gg T_D$.

\begin{figure}
\includegraphics[width=.49\textwidth]{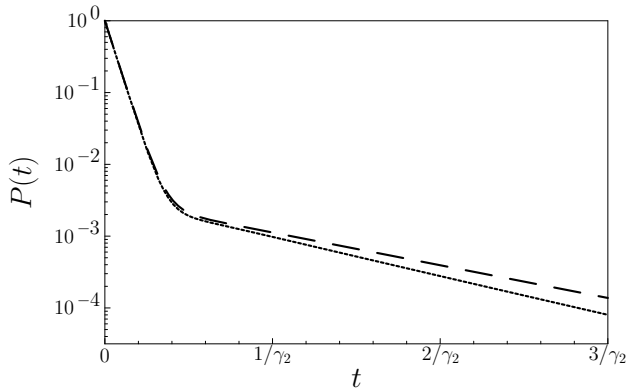}
\caption{Waiting-time distribution $P(t)$ as a function of the time interval between two subsequent quantum jumps events, in units of $1/\gamma_2$. The waiting-time distribution corresponding to the solid line has been evaluated numerically from Eq.~\eqref{eq:pnull}, whereas the dashed curve follows from the perturbative treatment in App.~\ref{app:tbd}. Two well separate time scales can be observed, from the short one follows the bright period $T_B$, the long one corresponds to the time scale of the dark period, $T_D$. For the considered parameters, they are in agreement with the expressions given in Eqs.~\eqref{eq:tbres} and~\eqref{eq:tdres}. The parameters are $\gamma_1=12\nu$, $\delta_1=6\nu$, $\Omega_1=2.5\nu$ and $\gamma_2=0.015\nu$, $\Omega_2=0.5\nu$, $\delta_2=0.87\nu$. The Lamb-Dicke parameters are taken to be  $\eta_1=\eta_2=0.05$, the angle of the laser wave vectors are $\phi_1=\pi/9$ and $\phi_2=0$, the detector is at angle $\psi=4\pi/5$ (not relevant for this plot). The mean phonon number at steady state is $\langle n\rangle \approx 0.3$.}\label{p0fig}
\end{figure}

In order to evaluate $T_B$ and $T_D$ we consider the waiting-time distribution $P(t)$, which is the probability that no photon emission occurs in the interval of time $t$, provided that a photon was detected at the instant $t=0$. The waiting-time distribution is defined as~\cite{Plenio98,Cohen86}
\begin{equation} P(t)=\Tr
\{e^{-{ i}\,H_{\rm eff}t/\hbar}\rho_{gg}
e^{{i}\,H^\dagger_{\rm eff}t/\hbar} \},
\label{eq:pnull}
\end{equation}
where
\begin{equation}
H_{\rm eff}=H-{ i\hbar}\sum_j\frac{\gamma_j}{2}\ket j\bra j
\label{eq:Heff}
\end{equation}
is a non-Hermitian operator, with $H$ given by Eq.~(\ref{eq:hamiltanH}), and $\rho_{gg}$ is the density matrix describing the state of the system, assuming that at $t=0$ the system is found in the internal ground state. In our case the density operator $\rho_{gg}$ corresponds to performing a measurement at time $t=0$ on the steady state density matrix, whose outcome is the projection onto state $\ket g$. Hence, in our formalism
$$\rho_{gg}=\ket g\bra g {\rm Tr}_{\rm I}\rho_{\rm st}=\ket g\bra g \mu_{\rm st}\,,$$ where $\mu_{\rm st}$ is the steady state density matrix, resulting from the laser cooling process (see App.~\ref{app:cooling}). Figure~\ref{p0fig} displays the waiting function $P(t)$ as a function of time for a specific choice of the parameters. One observes a fast decay of $P(t)$ for short times, which is followed by a slow decay with a smaller rate at larger times.
A time scale $\tau$ can be identified, which separates the two behaviors. From $P(t)$ we can estimate the average time scales~\cite{Plenio98}
\begin{alignat}{1}
T_{\rm B} &= \left[1-P(\tau)\right]^{-1}\left\{\tau
+ P(\tau)^{-1}\int_0^{\tau} dt' P(t')\right\}\label{eq:td},\\
T_{\rm D} &= \tau + P(\tau)^{-1}\int_{\tau}^\infty dt' P(t')
\label{eq:tb}
\end{alignat}
for the bright and dark periods, respectively. We calculate the two time scales perturbatively up to second order in the Lamb-Dicke parameter. The details of the calculation are reported in App.~\ref{app:tbd} and yield
\begin{alignat}{1}
T_{\rm B} &\approx \frac{1}{\eta_2^2\cos^2\phi_2}\frac{4\Gamma^2}{\Omega_2^2}T^{(0)}
\label{eq:tbres},\\
T_{\rm D} &\approx\frac{1}{\gamma_2},
\label{eq:tdres}
\end{alignat}
where
\begin{equation}
T^{(0)}=\left(\gamma_{1}\rho_{11}\right)^{-1}
\label{eq:t0}
\end{equation}
is the time scale of photon emission from state $\ket 1$, being $\rho_{11}$ the occupation of state $\ket 1$ at steady state and leading order in the Lamb-Dicke expansion, $\rho_{11}=\frac12 s/(s+1)$, with the saturation parameter~\cite{cohenbook}
\begin{equation}
s=\frac{\Omega_1^2/2}{\delta_1^2+\gamma_1^2/4}
\end{equation}
for the transition $\ket g\leftrightarrow\ket 1$.
The explicit form of the temperature-dependent parameter $\Gamma$ is  given in Eq.~\eqref{eq:gamma2}.
We assume small saturation $s\ll 1$ and $\gamma_2\ll s\gamma_1/2$, corresponding to the assumption that the rate $\gamma_1'=s\gamma_1/2$ of incoherent scattering on the transition $\ket g\leftrightarrow \ket 1$ is faster than the decay rate of state $\ket 2$. In this limit, we find
\begin{equation}
\label{eq:Gamma2}
\Gamma^2\approx
\frac{{\gamma'_1}^2}{4\langle n\rangle}.
\end{equation}
Using Eq.~\eqref{eq:Gamma2} and Eq.~\eqref{eq:t0}, for $s\ll 1$, in Eq.~\eqref{eq:tbres}, the scale for the bright periods takes on the form
\begin{equation}
T_B\approx \left(\eta_2^2\cos^2\phi_2\langle n\rangle \frac{\Omega_2^2}{\gamma'_1}\right)^{-1}\,,
\end{equation}
which corresponds to the time scale for scattering a photon along the transition $\ket g\leftrightarrow \ket 2$. Moreover, since scattering along this transition is exclusively due to the mechanical effects of light, the time scale of the bright periods becomes larger, the smaller are either the atomic spatial fluctuations about the trap center, since they scale with $\Delta x^2 \sim \xi^2\cos^2\phi_2\langle n\rangle $, or the smaller is the mechanical effect of the standing wave laser on the atom (which indeed vanishes when the laser propagates perpendicularly to the atomic motion, $\cos\phi_2=0$). In particular, when $\cos\phi_2=0$ there are no excitations of state $\ket 2$ and hence no quantum jumps are observed, corresponding to $T_B\to \infty$.

\subsection{Evaluation of the spectrum of resonance fluorescence}

The relation between the temporal behavior of the fluorescence signal, governed by the dark and bright periods, with the spectrum of resonance fluorescence was studied in Refs.~\cite{Plenio95,Plenio98} for an atom at rest. There, the authors found the occurrence of a sharp peak at the laser frequency, whose width scales with the inverse lifetime of the shelving period. We now extend these studies to the case of a trapped atom in the Lamb-Dicke regime, using the formalism first introduced in~\cite{Lindberg86,Cirac93} and then further developed in~\cite{Bienert04,Bienert06,BienertTorres07}. We evaluate the spectrum of resonance fluorescence by decomposing the spectrum in the eigenbasis of the Liouville operator, Eq.~\eqref{eq:fullmastereq} and evaluate the resulting expressions perturbatively in the Lamb-Dicke parameter. Such an approach allows us for discussing the relevant components of the spectrum belonging to the two transitions $\ket g\leftrightarrow\ket j$ individually.

We consider the spectrum of the fluorescence light scattered by the atom along the transition $\ket g\leftrightarrow \ket j$  and recorded in the far-field. The atom is assumed to be in the (dynamical) steady state of laser cooling. The component of the spectrum at frequency $\omega$ is proportional to the function
\begin{equation}
S^{(j)}(\omega) = {\rm Re}\int\limits_0^\infty dt\, e^{-i(\omega-\omega_{{\rm L}j})t}
\langle {D^{(j)}}^\dagger(t)D^{(j)}\rangle_{\rm st}
\label{eq:spec}
\end{equation}
with
\begin{alignat}{1}
D^{(j)} &= \ketbra gj \exp[- i k_j x \cos\psi],
\label{eq:dipop}
\end{alignat}
giving the positive-frequency component of the dipole operator, which describes atomic de-excitation and the recoil of the emitted photon which is measured by the detector \footnote{We assume that the transition
frequencies are widely separated such that there is no interference between the
spectral signals of each of the two transitions. This allows to treat the spectrum
of each transition separately.} at angle $\psi$. Note that $\langle \cdot\rangle_{\rm st}\equiv {\rm Tr}\{\cdot \varrho_{\rm st}\}$, where $\varrho_{\rm st}$ is the steady state defined by ${\mathcal L}\varrho_{\rm st}=0$. In Eq.~\eqref{eq:spec} we omit an overall
prefactor including the dipole radiation pattern of the transition.
In the following discussion we omit the superscript $j$ when discussing the general properties.

Using the quantum regression theorem, we can rewrite the expectation value in Eq.~(\ref{eq:spec}) as $\langle D^\dagger(t)D\rangle_{\rm st}={\rm Tr}\{D^\dagger e^{\cL t}D\varrho_{\rm st}\}$. Inserting the completeness relation~\eqref{eq:completp} between the operators, the power spectrum,
Eq.~\eqref{eq:spec}, takes the form
\begin{equation}
S(\omega) = {\rm Re}\sum_{\lambda}\frac{1}{i(\omega-\omega_{\rm L})-\lambda} F(\lambda)
\label{eq:specff}
\end{equation}
where
\begin{equation}
F(\lambda)={\rm Tr}\{{D}^\dagger\cP^\lambda D\varrho_{\rm st}\}
\label{eq:F}
\end{equation}
is a weight function of each spectral contribution. The explicit form of the spectral component can be determined using the Lamb-Dicke expansion of the Liouville operator and of the Dipole operator $D(x)$. In particular, in zero order in the Lamb-Dicke parameter the spectrum has the form
\begin{equation}
S_0(\omega) = {\rm Re}\sum_{\lambda_{\rm I}}
\frac{1}{i(\omega-\omega_{\rm L})-\lambda_{\rm I}}F_0(\lambda_{\rm I}),
\label{eq:spec0}
\end{equation}
with
$
F_0(\lambda_{\rm I}) = \Tr\{D_0^\dagger\cP_0 D_0\rho_{\rm st}\}
$
and $D_0=D(0)$. Equation~\eqref{eq:spec0} gives the spectrum of the bare atom at rest. It consists of the $\delta$-like elastic peak at the laser frequency, corresponding to the eigenvalue $\lambda_{\rm I}=0$, and of the inelastic components, corresponding to $\lambda_{\rm I}\neq0$.  For a two-level transition it corresponds to the spectrum evaluated by Mollow~\cite{Mollow69}.

The first non-vanishing contributions to the spectrum due to the mechanical effects of light are at second order in the Lamb-Dicke expansion, and we denote them by the term
\begin{equation}
S_2(\omega)={\rm Re}\sum_{\lambda}
\frac{1}{i(\omega-\omega_{\rm L})-\lambda}F_2(\lambda),
\label{eq:scndorderspec}
\end{equation}
with
\begin{equation}
F_2(\lambda) =\hspace{-4mm} \sum_{\alpha+\beta+\gamma+\delta=2}
\hspace{-4mm}
\Tr\{D^\dagger_\alpha \cP^\lambda_\beta D_\gamma\varrho^{\rm st}_\delta\},
\label{eq:internaltraces}
\end{equation}
and $D_n=\partial^n D(x)/\partial x^n|_{x=0}/n!$ is the corresponding coefficient of the expansion of the dipole operator. For later convenience we split the second order spectrum $S_2(\omega)$ into the correction to the elastic peak, corresponding to the contributions in second order in the Lamb-Dicke parameter at frequency $\omega=\omega_{{\rm L}}$ and here denoted by $S_{\rm el}(\omega)$, the contribution at the frequencies $\omega=\omega_{{\rm L}}\pm \nu$, corresponding to the motional sidebands of the elastic peak, given by the term
\begin{equation}
S_{\rm sb}(\omega) = {\rm Re}\!\!\!\!\!\!\!
\sum_{\genfrac{}{}{0pt}{2}{\lambda}{\lambda_{\rm I}=0,\lambda_{\rm E}=\pm i\nu}}
\!\!\!\!\!\!\!
\frac{1}{i(\omega-\omega_{\rm L})-\lambda}F_2(\lambda),
\end{equation}
and finally the contribution from inelastic scattering,
\begin{equation}
S_{\rm inel}(\omega) = {\rm Re}\sum_{\genfrac{}{}{0pt}{2}{\lambda}{\lambda_{\rm I}\neq 0}}
\frac{1}{i(\omega-\omega_{\rm L})-\lambda}F_2(\lambda),
\label{eq:specinel}
\end{equation}
such that $S_2(\omega)=S_{\rm el}(\omega)+S_{\rm sb}(\omega)+S_{\rm inel}(\omega)$. In general, the inelastic component at second-order in the Lamb-Dicke parameter is a small correction to the inelastic component at zero order.
Detailed examples on how to evaluate the spectra can be found in~\cite{Bienert04,Bienert06}.

The spectra are solved evaluating numerically the traces $F_0$ and $F_2$. Nevertheless, in some limit analytical expressions can be derived for the terms of interest to this work, namely, the ones related to the appearance to quantum jumps in the temporal behavior of the resonance fluorescence.

\subsection{Spectra of resonance fluorescence: Results}
\label{sec:resultsspec}

\begin{figure}
\includegraphics[width=7.5cm]{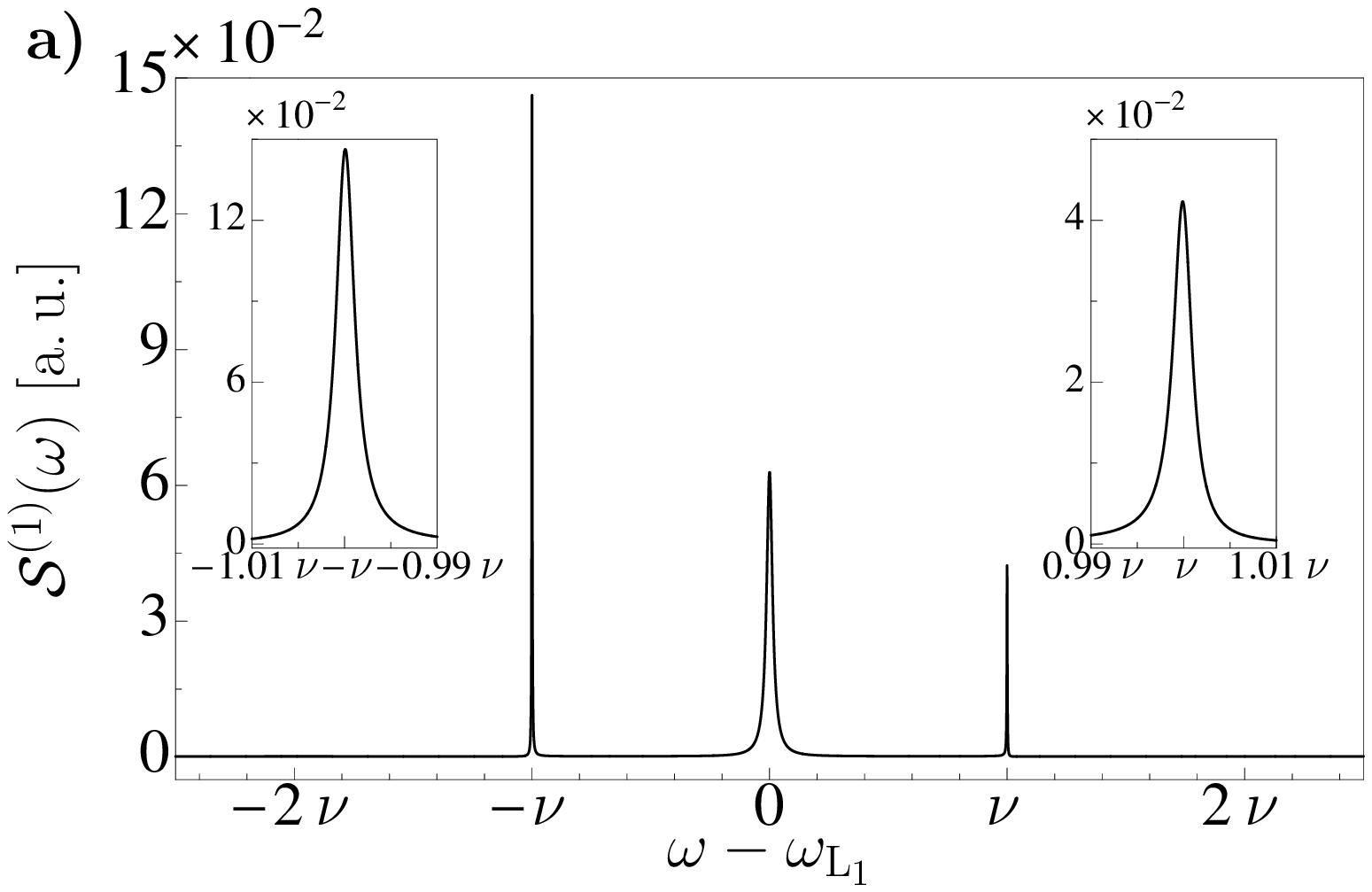}
\includegraphics[width=7.5cm]{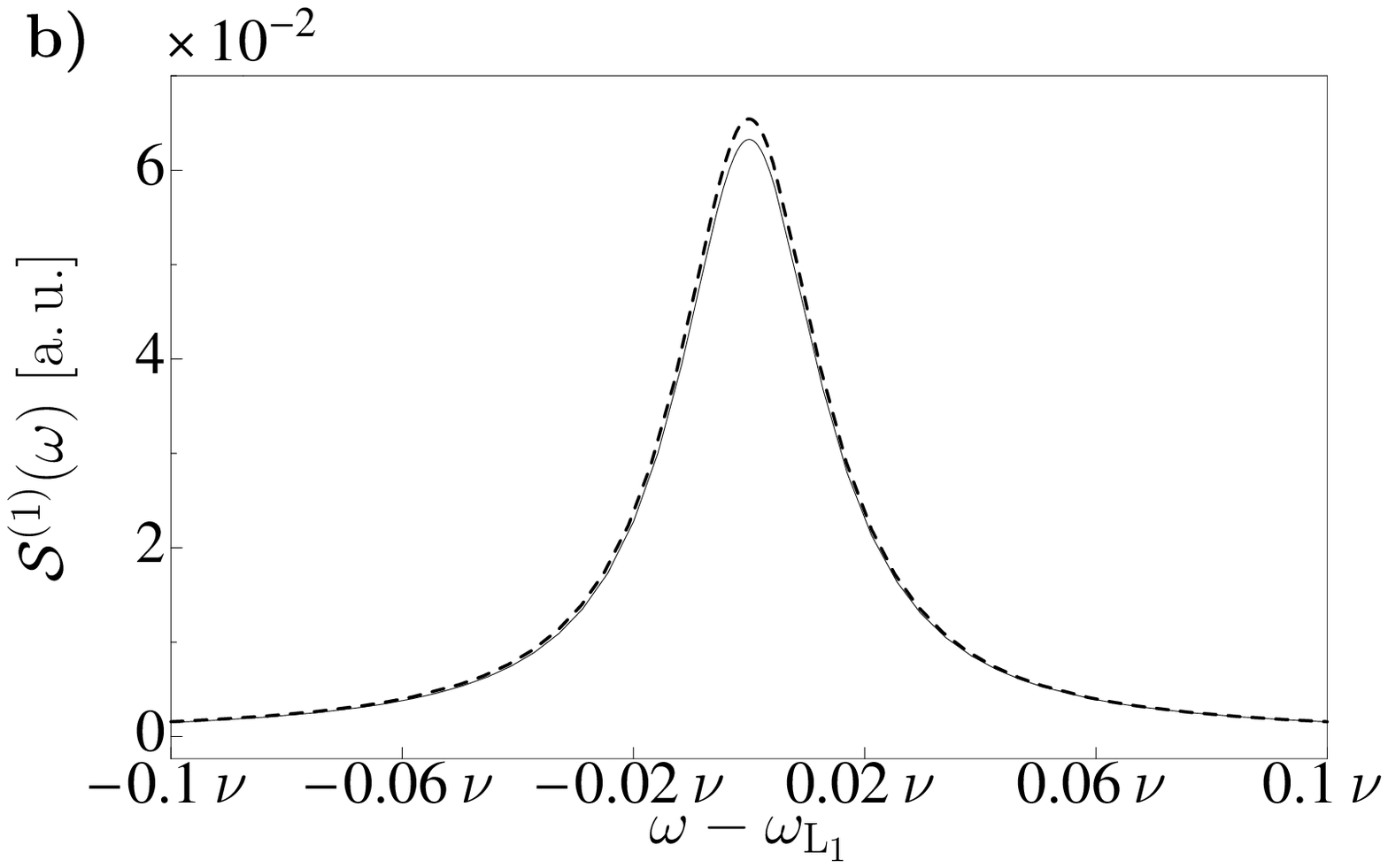}
\caption{\label{fig:spec1}
a) Spectrum of resonance fluorescence of the light emitted by state $\ket{1}$.
The inset shows the magnified sidebands of the elastic peak. The visible peaks are due to mechanical effects of light. b) Magnification of  the central narrow peak. The solid line represents the spectrum from Eqs.
(\ref{eq:spec0}) and (\ref{eq:scndorderspec}), while the dashed line corresponds to the approximation using the expressions (\ref{eq:peak}) and (\ref{eq:nps}). Same parameters as in Fig.~\ref{p0fig}.
}
\end{figure}

Figure~\ref{fig:spec1} a) displays the spectrum of resonance fluorescence of the light emitted from the atom along the transition $\ket g\leftrightarrow \ket 1$, where the elastic peak, at $\omega=\omega_{{\rm L}1}$, is not shown. Three peaks are visible: (i) the two sidebands of the elastic peak, corresponding to the photons emitted at frequency $\omega_{{\rm L}1}+\nu$ and $\omega_{{\rm L}1}-\nu$, and which are associated with the scattering processes which cool and heat, respectively, the center-of-mass motion by one phonon, and (ii) a central peak at $\omega=\omega_{{\rm L}1}$, which is due to the inelastically scattered photons. These three features are in second order in the Lamb-Dicke parameters, and are hence solely due to the mechanical effects of light. Further components, like for instance the inelastic components at zero order in the mechanical effects, are nonvisible and completely overtopped by the signals due to the mechanical effects. The sidebands of the elastic peak have the form already known from Doppler cooled atoms in the asymptotic stage of cooling, and have been object of several extensive studies~\cite{Lindberg86,Cirac93,Bienert04,Bienert06}. Here, we focus on the central peak, which is a novel feature with respect to the spectrum recorded by a two-level atom which undergoes Doppler cooling. An analytical form of the behavior of the spectrum around this point can be extracted from our theory. For $s\ll 1$ and assuming $\gamma_2$ to be the smallest decay rate, ($\gamma_2\ll s\gamma_1/2$) it has the form
\begin{equation}
S^{(1)}_{\rm inel}(\omega)\Bigl|_{\omega\simeq \omega_{{\rm L}1}} \approx \frac{\gamma_2}{(\omega-\omega_{{\rm L}1})^2+\gamma_2^2}\frac{T_D}{T_B}\frac{s}2,
\label{eq:s1inel}
\end{equation}
where the details of the calculations are reported in Appendix~\ref{app:evinttraces}. The analytical result in a more precise form, see Eqs.~\eqref{eq:peak} and 
\eqref{eq:nps}, is compared to the numerical one in Fig.~\ref{fig:spec1} b) and shows good agreement. This contribution is the one giving rise to the pedestal of the elastic peak due to the interruption of fluorescence, and was first described in Ref.~\cite{Plenio95} for the case of an atom at rest. This central peak can be understood as a
consequence of the random telegraphic-like form of the fluorescence signal~\cite{Plenio95,Plenio98}, being the appearance of dark periods, in which the atom does not emit photons, associated with the excitation of state $\ket 2$, and hence giving rise to a line broadening of the ground state $\ket g$. In our case, this dynamics is due to the mechanical effect of light, and indeed $T_B$ scales inversely proportional to the localization of the atoms about the trap center.

\begin{figure}
\includegraphics[width=7.5cm]{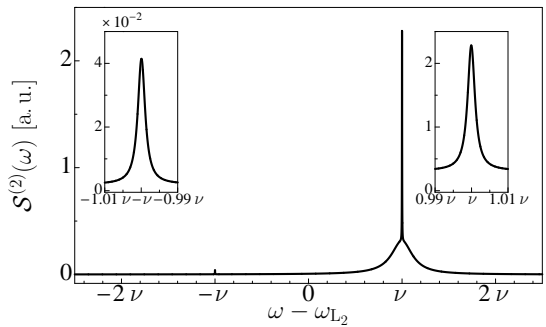}
\caption{
\label{fig:spec2}
Spectrum of resonance fluorescence of the light emitted by state $\ket{2}$.
The sidebands are magnified in the insets. Same parameters as in Fig.~\ref{p0fig}.
}
\end{figure}

Let us now analyze the corresponding features of the quantum jumps in the spectrum of resonance fluorescence of the light emitted along the transition $\ket g\leftrightarrow\ket 2$. Figure~\ref{fig:spec2} displays the spectrum derived by numerically evaluating the various terms to second order in the Lamb-Dicke expansion. As there is no excitation of state $\ket 2$ at lowest order, all contributions to the spectrum are at second order in $\eta$. We observe two features (also in this case the elastic peak is not reported): the blue sideband of the elastic peak with one large pedestal. This characteristic is independent of the angle of observation. The other sideband, corresponding to the transitions heating the atom, is hardly visible. From the magnifications in the insets, one can see that its height is two orders of magnitude smaller than the height of the blue sideband. 

We first discuss the asymmetry between the two sidebands. The spectra are evaluated at steady state of laser cooling where detailed balance of the heating and cooling rates between the phononic populations is satisfied. These rates are given by the transition rates connected to the scattering processes of laser cooling, multiplied by the population of the vibrational levels. For a single atomic dipole involved in the laser cooling, the strengths of the red and blue sidebands integrated over all directions of emission, is essentially given by these heating and cooling rates, which are equal. For multiple atomic transitions participating in the cooling, the motional sidebands reflect only the scattering processes belonging to the considered dipole, and thus their strengths can differ. In the competition of the two cooling mechanisms, the sideband cooling of the standing wave laser coupling transition $\ket g\leftrightarrow\ket 2$ dominates for the parameters used here. As a result, in the spectrum of this transition we observe a strongly pronounced blue sideband and an almost completely suppressed red sideband. This asymmetry is balanced out by the other situation in the angle-integrated spectrum of the Doppler cooled transition. Our intuitive explanation is also supported by an analytical calculation, which can be done at leading order in the saturation parameter $s$, and is reported in Appendix~\ref{app:evinttraces}. The situation encountered here is reminiscent to the case of cavity cooling of a trapped atom \cite{BienertTorres07}, where also two decay channels -- the spontaneous emission of the atom and the cavity loss -- exist.

The pedestal appearing at the location of the blue sideband is an inelastic contribution to the spectrum, and arises from the broadening of the blue sideband transition due to the coupling with state $\ket 1$. Its central frequency is determined by the choice of the parameters of the laser coupling to state $\ket 2$. An analytical calculation shows that it has the form
\begin{equation}
S_{\rm inel}^{(2)}(\omega)\Bigl|_{\omega\simeq \omega_{{\rm L}2}+\nu}\approx \frac{T_D}{T_B}\frac{\gamma_1'/2}{\gamma_1'^{2}/4+(\omega_{{\rm L}2}+\nu-\omega)^2}
\label{eq:s2inel}
\end{equation}
where $\gamma_1'=s\gamma_1/2$ and which has been derived assuming $s\ll 1$.  In the regime discussed in Sec.~\ref{sec:qj}, in which quantum jumps are observable, it is hence dominant. Interestingly, its height is proportional to the temperature of the atom, as one can check using the explicit form of the time scales $T_D$ and $T_B$, Eqs.~(\ref{eq:tdres}) and~(\ref{eq:tbres}), and is also proportional to the degree of localization of the atom around the trap center, $\Delta x^2\sim\xi^2\langle n\rangle$.

We finally discuss the dependence of the spectral signals on the angles $\phi_1$ and $\phi_2$ of the incident lasers and on the detector position $\psi$. The $\psi$-dependence enters both in the dipole pattern of the emitted radiation, which is an overall factor, and in the individual spectral components. We focus on the latter ones. 
In the limit considered here, the motional sidebands at $\omega\approx\omega_{\rm L1}\pm\nu$ are the only signals depending on $\psi$. Their angle dependence is discussed in Ref.~\cite{Cirac93} and emerges from the quantum interference between photon scattering processes leading to an exchange of a vibrational excitation. For $\phi_1=\pi/2$, {\it i.e.} when the driving laser is perpendicular to the axis of motion, the motional sidebands are only due to diffusive processes, resulting in a $\cos^2\psi$-dependence.
The strength of the central peak in the spectrum of transition $\ket
g\leftrightarrow \ket 1$ is proportional to $T_B^{-1}\propto \cos^2\phi_2$,
hence it depends on the orientation of the second laser driving the other
transition $\ket g\leftrightarrow \ket 2$. Similarly, our results show that
$S^{(2)}(\omega)\propto \cos^2\phi_2$: All the spectral components of this
transition are due to the mechanical effects of the second laser, and
disappear for $\phi_2= \pi/2$. Moreover, the laser angles determine the mean
vibrational excitation number $\langle n\rangle$, Eq.~ \eqref{eq:meann}.
This quantity additionally scales the components of the spectra which are
related to mechanical effects, as shown in App.~\ref{app:evinttraces}.
The spectral features connected with the quantum jumps are optimally
observed when the wave vector of the standing wave is aligned with the
atomic motion ($\phi_2=0$) and the detector angle $\psi$ is at the maximum
of the dipole emission pattern. In the regime considered here, the orientation of the running wave laser is less essential, best visibility we found for illumination along the axis of motion ($\phi_1 = 0$).

\section{Discussion and conclusions}
\label{Sec:5}

We have theoretically studied the resonance fluorescence from a confined atom in a setup, where one could observe quantum jumps which are solely due to the mechanical effects of light. In the limit in which the lasers driving the atom also cool the motion, we have derived the time scales of the bright and dark periods of the emitted intensity and the corresponding features in the spectrum of resonance fluorescence. The latter is a pedestal appearing in correspondence of the elastic peak for the radiation emitted along the bright transition, whose height is proportional to the temperature and degree of localization of the atom about the trap center.

We now discuss the experimental feasibility of this scheme. The setup we have considered assumes a $V$-configuration of levels, where one of the excited state is metastable. The atom is Doppler cooled along the optically allowed transition, while the transition which is metastable is driven by a standing-wave laser. In this work we assumed that the linewidth of this transition is much narrower than the trap frequency, so that the corresponding laser essentially drives the red sideband. This is however not an essential requirement. The important condition is that the linewidth of the metastable state is significantly smaller than the rate of photon scattering along the allowed, optical transition. A situation like the one here considered could be encountered in $^{40}$Ca$^+$ ion, where the $V$-transition could be constituted by dipolar transition $S_{1/2}\to P_{1/2}$, where the atom is Doppler cooled, and quadrupole transition $S_{1/2}\to D_{5/2}$, which is driven by a standing wave field, like the one of an optical cavity, as realized in~\cite{Mundt}. Since state $D_{5/2}$ has a lifetime of the order of a second, this could be quenched by coupling to $P_{3/2}$, thereby not affecting the quantum jump scheme.

While the motivation of this work is mainly to investigate some fundamental properties of quantum jumps in resonance fluorescence, one could consider which outlooks on quantum engineering this work may offer. Beyond the possibility to extract temperature and localization of an atom, one could consider a setup in which the quantum jump corresponds to the transition to a nonclassical state of the motion. In this case, this would allow one to post-selecting the quantum state, in the spirit of the proposal in Ref.~\cite{Almut,Metz2007a,Metz2007b}.

\begin{acknowledgments}
This work was partly supported by the European Commission (EMALI, MRTN-CT-2006-035369; SCALA, Contract
No.\ 015714), by the Spanish Ministerio de Ciencia y Innovaci\'on (Consolider Ingenio 2010 "QOIT", CSD2006-00019; QNLP, FIS2007-66944; Ramon-y-Cajal and Juan-de-la-Cierva programs), and by project IN114310 by PAPIIT, Universidad Nacional Aut\'onoma de M\'exico. G.M. acknowledges support by the German Research Council (Heisenberg professorship). M.B. is indebted to the Alexander von Humboldt foundation for their support.
\end{acknowledgments}
\appendix

\section{Eigensystem of $\cL_{\rm I}$}
\label{app:esintern}
In this appendix we classify the left and right side eigenelements
$\check\rho_{\rm I}^{\lambda_{\rm I}}$ and
$\hat\rho_{\rm I}^{\lambda_{\rm I}}$ of the Liouville operator for the
internal motion $\cL_{\rm I}$, Eq.~\eqref{eq:Li}, obeying
\begin{equation}
\cL_{\rm I}\hat\rho_{\rm I}^{\lambda_{\rm I}}=
\lambda_{\rm I}\hat\rho_{\rm I}^{\lambda_{\rm I}} ,
\quad \check\rho_{\rm I}^{\lambda_{\rm I}}\cL_{\rm I}=
\lambda_{\rm I} \check\rho_{\rm I}^{\lambda_{\rm I}}.
\label{eq:inteel}
\end{equation}
A part of the eigensystem of $\cL_{\rm I}$ can be most
easily calculated using the right and left eigenstates
\begin{alignat}{1}
\ket{\pm}&=\left(\omega_{\pm}\ket 1+
\frac{\Omega_1}2\ket g\right)/\sqrt{\omega_{\pm}^2+\Omega_1^2/4}\\
\obra{\pm}&=\left(\omega_{\pm}\bra 1+
\frac{\Omega_1}2\bra g\right)/\sqrt{\omega_{\pm}^2+\Omega_1^2/4}
\label{eq:effestate}
\end{alignat}
and $\ket 2$
of the effective Hamiltonian 
\begin{equation}
H_{\rm int}^{(\rm eff)} = H_{\rm int} +
V(0)-i\hbar\sum_{j=1,2}\frac{\gamma_j}{2}\ketbra jj
\label{eq:hinteff}
\end{equation}
with eigenfrequencies
\begin{equation}
\omega_\pm = \frac{1}{2}\left(\delta_1-i\frac{\gamma_1}{2}\right)
\pm\frac{1}{2}
\sqrt{\left(\delta_1-i\frac{\gamma_1}2\right)^2+\Omega_1^2}
\label{eq:effefreq}
\end{equation}
and $\omega_2 = \delta_2-i\frac{\gamma_2}{2}$. By using
$\ketbra 2{\pm}$ and $\ket 2\obra{\pm}$ together with their
Hermitian conjugates as an ansatz, one finds
\begin{alignat}{3}
\check\rho_{\rm I}^{\lambda_{\rm I}} &= \ket2 \obra{\pm},
\quad&\hat\rho_{\rm I}^{\lambda_{\rm I}} &= \ketbra {\pm}2\quad
& \text{to}\quad \lambda_{1,\pm} = (\omega_\pm-\omega_2^\ast)/i \\
\check\rho_{\rm I}^{\lambda_{\rm I}} &= \oket{\pm}\bra2, \quad
&\hat\rho_{\rm I}^{\lambda_{\rm I}} &= \ketbra 2{\pm}\quad
& \text{to}\quad \lambda_{2,\pm} = (\omega_2-\omega_\pm^\ast)/i.
\end{alignat}
Apart from these four elements which describe the transition between
$\ket 2$ and the dressed $\ket g\leftrightarrow \ket 1$ transition,
we further have
\begin{alignat}{1}
\check\rho_{\rm I}^{\lambda_{\rightsquigarrow}} = \ketbra 22,
\quad \hat\rho_{\rm I}^{\lambda_{\rightsquigarrow}}=
\ketbra 22 - \frac{1}{\Upsilon+2|\varsigma|^2}\Big( |\varsigma|^2\ketbra 11
\nonumber\\
+[\Upsilon+|\varsigma|^2]\ketbra gg  -
\left\{\Upsilon \varsigma^\ast \ketbra 1g +{\rm H.c.}\right\}\Big)
\end{alignat}
to $\lambda_{\rightsquigarrow}=-\gamma_2$
with $\Upsilon=\frac{\gamma_1-\gamma_2}{\gamma_1-2\gamma_2}$ and
$\varsigma=\frac{\Omega_1}{2\delta_1+i(\gamma_1-2\gamma_2)}$ describing
the decay of population in $\ket 2$. The remaining four eigenelements are
those of the two-level subsystem $\ket g\leftrightarrow \ket 1$ with Rabi
frequency $\Omega_1$, detuning $\delta_1$ and decay rate $\gamma_1$. These
are
\begin{alignat}{3}
\check\rho_{\rm I}^{\lambda_{\rm I}} &= 1,
\quad&\hat\rho_{\rm I}^{\lambda_{\rm I}} &= \rho_{\rm st}
&\quad \text{to}\quad \lambda_{\rm st} = 0
\end{alignat}
belonging to the internal steady state
\begin{alignat}{1}
\rho_{\rm st} = \frac1{\mathcal N}_{\rm st}
\Big(&\Omega_1^2\ketbra11+({\mathcal N}_{\rm st}-\Omega_1^2)\ketbra gg
\nonumber\\
-&\left\{\Omega_1 [2\delta_1+i\gamma_1]\ketbra 1g +
{\rm H.c.}\right\}\Big)
\label{eq:intstst}
\end{alignat}
where
\begin{equation}
{\mathcal N}_{\rm st}=\gamma_1^2+4\delta_1^2+2\Omega_1^2,
\label{eq:Nst}
\end{equation}
and the three eigenvalues $\lambda_{\rm TLS}^0$, $\lambda_{\rm TLS}^\pm$.
From the latter ones, $\lambda_{\rm TLS}^0$ is real, the other two form
a complex conjugate pair at saturative driving. They represent the possible
transitions between the states of the dressed two-level subsystem and
correspond to its Mollow-triplet. For $\delta_1=0$ simple analytic expressions
for these eigenvalues and eigenelements can be found \cite{Jakob03}.

\section{Cooling}\label{app:cooling}
The atomic motion is cooled by the mechanical effects of the two lasers.
The motional state at the final stage of the cooling dynamics can be estimated
in perturbation theory~\cite{Cirac92,Eschner03}.
At lowest relevant order in the Lamb-Dicke parameter, the dynamics of the
population $p_n={\rm Tr}\pg{\kb{n}{n}\varrho}$ in the vibrational state $\ket n$
is described by the rate equation
\begin{alignat}{1}
\dot p_n=&(n+1)(A_{1-}+A_{2-})p_{n+1}+n(A_{1+}+A_{2+})p_{n-1}\\
&-\pq{(n+1)(A_{1+}+A_{2+})+n(A_{1-}+A_{2-})}p_n
\label{eq:rateeq}
\end{alignat}
where $(n+1)(A_{1+}+A_{2+})$ and $n(A_{1-}+A_{2-})$ are the rates for population
transfer between the levels $n$ and the levels $n+1$ and $n-1$ respectively.
The two terms $A_{1\pm}$ and $A_{2\pm}$ entering these rates account for the
mechanical effects of the lasers coupled to the transitions
$\ket{g}\leftrightarrow\ket{1}$ and $\ket{g}\leftrightarrow\ket{2}$.
They are given by
\begin{eqnarray}
  A_{1\pm}&=&2{\rm Re}\pg{s_1(\mp\nu)+D}\nonumber\\
  A_{2\pm}&=&2{\rm Re}\pg{s_2(\mp\nu)}
\end{eqnarray}
where
\begin{eqnarray}
s_j(\nu)&=&\frac{\xi^2}{\hbar^2}\int_0^\infty d\tau{\rm e}^{{\rm i}\nu\tau}{\rm Tr}
\pg{V_1^{(j)}{\rm e}^{{\cal L}_I \tau}V_1^{(j)}\rho_{\rm st}}\nonumber\\
D&=&\eta_1^2{\cal W}_2\frac{\gamma_1}{2}{\rm Tr}\pg{\kb{1}{1}\rho_{\rm st}}
\end{eqnarray}
with
\begin{eqnarray}
V_1^{(1)}&=&{\rm i}
\hbar\frac{\Omega_1}{2}k_1\cos\phi_1\pq{\kb{1}{g}-\kb{g}{1}}
\nonumber\\
V_1^{(2)}&=&-\hbar\frac{\Omega_2}{2}k_2\cos\phi_2\pq{\kb{2}{g}+\kb{g}{2}}
\end{eqnarray}
and $\rho_{\rm st}$ being the internal steady state from Eq.~\eqref{eq:intstst}.
The motional state at the final stage of cooling 
\begin{equation}
  \mu_{\rm st} = \frac{1}{1+\langle n\rangle}\left(\frac{\langle n\rangle}{1+\langle n\rangle}\right)^{a^\dagger a}
\end{equation}
follows from the rate equation~\eqref{eq:rateeq} for $\dot p_n=0$. The average number of vibrational excitations can be expressed in the form
\begin{eqnarray}\label{eq:finaln}
\av{n}=\frac{n_1W_1+n_2 W_2}{W}
\label{eq:meann}
\end{eqnarray}
where $n_j=A_{j+}/(A_{j-}-A_{j+})$ with $j=1,2$ is the excitation number corresponding to the situation in which only the transition $\ket g\leftrightarrow\ket j$ participates in the cooling process, and $W_j=A_{j-}-A_{j+}$ are the corresponding cooling rates. The total cooling rate which takes into account both transitions is given by
\begin{equation}
W=W_1+W_2.
\label{eq:coolingrate}
\end{equation}
 The steady state of the motional dynamics  results from the competition of
 the mechanical  effects of the photons scattered along the two transitions.

Simple expressions for $n_j$ and $W_j$ can be found when the parameters are set for
Doppler cooling on the transition $\ke{1}\leftrightarrow\ke{g}$ and sideband cooling
on the transition $\ke{2}\leftrightarrow\ke{g}$, i.e. when
$\gamma_1\gg\nu,\Omega_1,\Omega_2\gg\gamma_2$, $\delta_1=\gamma_1/2$ and
$\delta_2={\rm Re}\omega_-+\nu$.
The terms $n_1$ and $W_1$ are equal to the standard results of laser cooling
of a trapped two level atom and are given by
\begin{eqnarray}
n_1&\approx&\frac{\gamma_1}{4\nu}\pt{1+\frac{{\cal W}_2}{\cos^2\phi_1}}-\frac{1}{2}
\nonumber\\
W_1&\approx&2\eta_1^2\frac{\nu\Omega_1^2\cos^2\phi_1}{\gamma_1^2}.
\end{eqnarray}
The cooling on the transition $\ke{2}\leftrightarrow \ke{g}$ is influenced by the
dynamics on the other transition which is relevant at zero order, see
App.~\ref{app:esintern}.
In particular, the corresponding terms $n_2$ and $W_2$, are found by taking into account
the light shift ${\rm Re}\omega_-\simeq-\frac{\Omega_1^2}{4\gamma_1}$ of the ground
state due to the coupling to the first laser.
Together with the light shift the ground state acquire a finite line width
$\gamma_-\simeq\frac{\Omega_1^2}{2\gamma_1}$ and the corresponding results can be
expressed as
\begin{eqnarray}
n_2&\approx&\frac{\pt{\gamma_2+\gamma_-}^2}{16\nu^2}\nonumber\\
W_2&\approx&\frac{\eta_2^2\Omega_2^2\cos^2\phi_2}{\gamma_2+\gamma_-}.
\end{eqnarray}
\begin{figure}
\includegraphics[width=7.5cm]{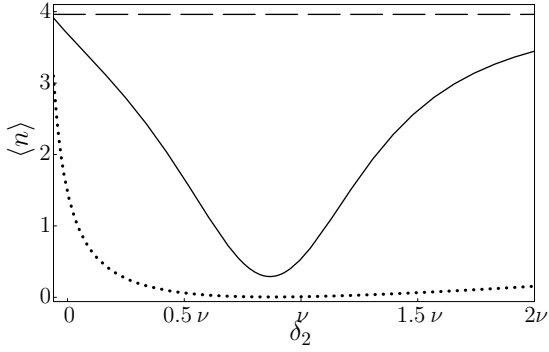}
\caption{\label{fig:temp}
Mean phonon number
$\langle n \rangle$ for the different cooling schemes
as function of $\delta_2$. The
dashed line shows the
resulting phonon number $n_1$ of the Doppler cooling only, while
the dotted line corresponds to the mean phonon number if
there was only sideband cooling at the standing-wave transition.
The solid line shows the mean phonon number from Eq.~\eqref{eq:finaln}
resulting from the competition between the two cooling
mechanisms. Same parameters as in Fig.~\ref{p0fig}, where $\delta_2$ was chosen to optimize the phononic excitation number $\langle n\rangle$.
}
\end{figure}

\section{Evaluation of the bright and dark time scales}
\label{app:tbd}
In order to find expressions for the bright and dark periods, we start from Eq.~\eqref{eq:pnull}, rewrite
$P(t) = \sum_m \bra m\mu_{\rm st}\ket m P_m(t)$
in the basis of the vibrational eigenstates $\ket{m}$ and expand
\begin{alignat}{1}
&P_m(t) = \sum_{n, n'} \Big[ e^{-i \Omega_{nn'} t}
\braket{\chi_{2,n'}}{\chi_{2n}}\obraket{\chi_{2n}}{m}
\ket{g}\bra{g}\braoket{m}{\chi_{2,n'}} \nonumber\\
&+ \sum_{\sigma, \sigma'=\pm} e^{-i \omega_{\sigma \sigma'nn'} t}
\braket{\chi_{\sigma',n''}}{\chi_{\sigma, n}}\obraket{\chi_{\sigma n}}{m}
\ket{g}\bra{g}\braoket{m}{\chi_{\sigma'n'}}\Big]
\label{eq:p0m}
\end{alignat}
in terms of the right eigenstates $\ket{\chi_{\pm, n}}$, $\ket{\chi_{2,n}}$
and left eigenstates $\obra{\chi_{\pm, n}}$, $\obra{\chi_{2,n}}$ of the
effective Hamiltonian $H_{\rm eff}$, Eq.~\eqref{eq:Heff},
with eigenvalues $\hbar\omega_{\pm n}$, $\hbar\omega_{2n}$.
We also defined the slowly decaying rates
$\Omega_{nn'}=\omega_{2n}-\omega^\ast_{2n'}$ and the fast decaying rates
$\omega_{\sigma \sigma'nn'}=\omega_{\sigma n}-\omega^\ast_{\sigma'n'}$ as abbreviations.
In the last step we have neglected the mixed terms which include rapid
and slowly decaying frequencies: They are only small corrections to the
second line in Eq.~\eqref{eq:p0m}.

To proceed, we evaluate Eq.~\eqref{eq:p0m} perturbatively in the Lamb-Dicke
parameters $\eta_j$, Eq.~(\ref{eq:smalleta}). To this end, we expand the
effective Hamiltonian,
\begin{equation}
H_{\rm eff} = H_0^{\rm (eff)} + H_1^{\rm (eff)} + \dots
\label{eq:heff}
\end{equation}
its eigenfrequencies,
\begin{alignat}{1}
\omega_{\pm n} &= \omega^{(\pm n)}_{0} + \omega^{(\pm n)}_{1} + \dots\\
\omega_{2n} &= \omega^{(2,n)}_{0} + \omega^{(2,n)}_{1} + \dots
\end{alignat}
and the right and left eigenvectors
\begin{alignat}{1}
\ket{\chi_{2,n}}&=\ket{\chi_{2, n}}_0+\ket{\chi_{2, n}}_1+\dots\\
\obra{\chi_{2,n}}&= _{0\!\!}\obra{\chi_{2, n}}+ _{1\!\!}\obra{\chi_{2, n}}
+\dots\\
\ket{\chi_ {\pm, n}}&=\dots
\end{alignat}
 in increasing order in $\eta_j$. In lowest order, $H_0^{\rm (eff)}$ is
 a sum of the external part $H_{\rm ext}$ and the internal part
$H^{(\rm eff)}_{\rm int}$ from
Eq.~\eqref{eq:hinteff} 
whose right and left eigenstates $\ket{\pm}, \ket2$ and
$\obra{\pm}, \bra2$ with their eigenvalues $\omega_\pm$ and
$\omega_2$ are listed in App.~\ref{app:esintern},
Eqs.~\eqref{eq:effestate}-\eqref{eq:effefreq}. Moreover, we have
\begin{alignat}{1}
\ket{\chi_{2,n}}_0 = \ket 2\ket n, \quad
\ket{\chi_{\pm,n}}_0 = \ket \pm\ket n
\end{alignat}
and analogous expressions for the left eigenstates.
The higher order corrections we find using standard techniques
of perturbation theory. We now return to Eq.~\eqref{eq:p0m} and
evaluate
\begin{alignat}{1}
P_m(t) =& \sum_{n'} e^{-\gamma_2 t}| _{1\!}
\obraket{\chi_{2,n'}}{gn}|^2\nonumber\\
&+\sum_{\sigma,\sigma'}
e^{-i (\omega_\sigma-\omega_{\sigma'}^\ast)t}\braket{\sigma}{\sigma'}
\obraket{\sigma}{g}\braket{g}{\sigma'l}.
\label{eq:p0mres}
\end{alignat}
up to second order in the Lamb-Dicke parameter. It turns out that
we only need to calculate the first order correction
\begin{alignat}{1}
 _{1\!}\obra{\chi_{2,n}} =
 \eta_2\cos\phi_2&\left(\sqrt{n}\bra{n-1}+
 \sqrt{n+1}\bra{n+1}\right)\nonumber\\
&\times\bra{g} \frac{\hbar\Omega_2/2}{\hbar\omega_0^{(2,n)}-{H_0^{\rm (eff)}}}.
\label{eq:secordcorrbra}
\end{alignat}
Using expressions \eqref{eq:p0mres} and \eqref{eq:secordcorrbra} in
Eqs.~\eqref{eq:td} and \eqref{eq:tb} finally yields the expressions $T_B$ and $T_D$ from Eqs.~\eqref{eq:tbres} and \eqref{eq:tdres} when taking into account the separation of the time scales
$\abs{{\rm Im}\omega_\pm}\gg\frac{1}{\tau}\gg\abs{{\rm Im}\omega_2}$. In Eq.~\eqref{eq:tbres} we use the abbreviations
\begin{alignat}{1}
T^{(0)}&=-i\sum_{\sigma, \sigma'=\pm}
\frac{\braket{\sigma}{\sigma'}\obraket{\sigma}{g}\braoket{g}{\sigma'}}{(\omega_\sigma-\omega_{\sigma'}^\ast)}
=\frac{1}{\gamma_{1}\bra1\rho_0^{\rm st}\ket 1}\\
\Gamma^{-2} &= \sum_{\ell=\pm 1}(\langle n\rangle+\delta_{\ell,1})\left|\sum_{\sigma=\pm}
\frac{\braket{g}{\sigma}\obraket{\sigma}{g}}{\omega_2-\omega_\sigma+\ell\nu}\right|^2
\label{eq:gamma2}.
\end{alignat}
The time $T^{(0)}$ is the average waiting time between two photon emissions of the atom at zero order in the Lamb-Dicke expansion and $\Gamma$ determines the temperature dependence of the bright period. An expansion in the saturation parameter $s$ of this quantity gives
\begin{alignat}{1}\label{eq:Gamma2apps}
\Gamma^{-2}\approx
\frac{16\langle n\rangle}{\gamma_1^2 s^2}\left(
1+s+\frac{4}{s}\frac{\gamma_2}{\gamma_1}
\right).
\end{alignat}

\section{Evaluation of the spectrum of resonance fluorescence}
\label{app:evinttraces}

In this appendix we sketch the derivation of the analytical expressions for several components of the spectrum of resonance fluorescence. The central peak of transition $\ket g\leftrightarrow\ket 1$ can be calculated starting from Eq.~\eqref{eq:specinel} with $\lambda_{\rightsquigarrow}=-\gamma_2$ and
$\lambda_{\rm E} = 0$. We find
\begin{equation}
  S^{(1)}_{\rm cp}(\omega) = {\rm Re}\frac{1}{i(\omega-\omega_{{\rm L}1})-
\lambda_{\rightsquigarrow}}F^{(1)}_{\rm cp}
\label{eq:peak}
\end{equation}
where only
\begin{equation}
F^{(1)}_{\rm cp}=\left.
\Tr\{D_0^\dagger\cP^\lambda_2 D_0\rho_0^{\rm st}\}+
\Tr\{D_0^\dagger\cP^\lambda_1 D_0\rho_1^{\rm st}\}\right|_{\lambda=\lambda_{\rightsquigarrow}}
\label{eq:conttonp}
\end{equation}
of the ten possible second order combinations
of the internal traces $F_2(\lambda)$, Eq.~\eqref{eq:internaltraces}, contribute. The required techniques for evaluating the remaining traces can be found, {\it e.g.} in
Refs.~\cite{Bienert06, Bienert04}. It turns out, that
$\Tr\{D_0^\dagger\cP_2 D_0\rho_0^{\rm st}\}\propto\gamma_2^{-1}$ dominates.
Constraining ourselves on this term and exploiting
$\gamma_2\ll\gamma_1$, we find
\begin{equation}
F^{(1)}_{\rm cp}
\approx \eta_2^2 \cos^2\phi_2
\frac{\Omega_2^2}{2\gamma_2}|\langle 1|\rho_{\rm st}|
g\rangle |^2 P(\langle n\rangle)
\end{equation}
\label{eq:Fnp}
where
\begin{equation}
P(\langle n\rangle) = {\rm Re}\sum_{\sigma = \pm}
\langle g|\sigma\rangle\obra{\sigma}\rho_{\rm st}\ket g\frac{(2\langle n\rangle + 1)
(\gamma_2+\lambda_{1\sigma})-i\nu}{(\gamma_2+\lambda_{1\sigma})^2+\nu^2}.
\end{equation}
Here, $\oket{\sigma}$ and $\ket\sigma$ denote the left and right eigenstates of the
effective Hamiltonian, \eqref{eq:hinteff},
and $\lambda_{j\pm}$ are eigenvalues of the Liouville operator ${\mathcal L}_{\rm I}$
discussed in App.~\ref{app:esintern}. 
Taking into account the conditions for Doppler cooling and sideband cooling at the corresponding atomic transition, we obtain for the peak's weight function
\begin{equation}
F^{(1)}_{\rm cp}
\approx \eta_2^2 \cos^2\phi_2\langle n\rangle
\frac{\Omega_2^2}{\gamma_1\gamma_2}\left(1-2s+\frac{2}{s}\frac{\gamma_2}{\gamma_1}\right)
\label{eq:nps}
\end{equation}
up to first order of saturation parameter $s$.

We now turn to transition $\ket g\leftrightarrow\ket 2$. Here, three spectral signals of second order are observable: Two motional sidebands of the elastic peak,
\begin{equation}
  S^{(2)}_{\rm sb}(\omega) = {\rm Re}\frac{1}{i (\omega-\omega_{{\rm L}2})\pm i\nu+
\gamma_{\rm sb}}F^{(2)}_{\rm sb}(\pm i\nu)
\label{eq:sb2}
\end{equation}
and an inelastic contribution,
\begin{equation}
S^{(2)}_{\rm inel}(\omega) = {\rm Re}\frac{1}{i(\omega-\omega_{{\rm L}2})+{\lambda_{1-}}}
F^{(2)}_{\rm inel},
\label{eq:peak2}
\end{equation}
belonging to the eigenvalue $\lambda_{1-}$ of ${\mathcal L}_{\rm I}$. In Eq.~\eqref{eq:sb2} we introduced the width of the
motional sidebands, $\gamma_{\rm sb}=W/2$, given by the real part
of the second order correction $\lambda_2$ and corresponding to the cooling
rate $W$, Eq.~\eqref{eq:coolingrate}, of the laser cooling process
\cite{Cirac92}.

In order to characterize the form of the spectral signals, we need to
evaluate the weighting terms $F^{(2)}_{\rm sb}(\pm i \nu)$ and
$F^{(2)}_{\rm inel}$. For the motional sidebands,
only the term $\Tr\{D_0^\dagger\cP_1 D_0\rho_1^{\rm st}\}$ contributes
and yields
\begin{alignat}{1}
F^{(2)}_{\rm sb}(i\ell\nu) &= \eta_2^2\cos^2\phi_2
\frac{\Omega_2^2(\langle n\rangle + \delta_{\ell,-1})}{4}\times\nonumber\\
&\left|\sum_{\sigma=\pm}\frac{\langle g\ket{\sigma}\obra{\sigma}
\rho_{\rm st}\ket{g}}{\lambda_{1\sigma}-\lambda_{\rm E}}\right|^2.
\label{eq:fsb2}
\end{alignat}
The real expression
Eq.~\eqref{eq:fsb2} indicates that the motional sidebands are pure
Lorentzians and independent from the detector position angle $\psi$.
An expansion in the saturation parameter $s$ under the same conditions as
for Eq.~\eqref{eq:nps}, reveals
\begin{alignat}{1}
F^{(2)}_{\rm sb}(-i\nu) &=
\eta_2^2\cos^2 \phi_2
\frac{\Omega_2^2(\langle n\rangle +1)}{16\nu^2}\left(1-s\right)\\
F^{(2)}_{\rm sb}(i\nu) &=
\eta_2^2\cos^2\phi_2
\frac{4\Omega_2^2\langle n\rangle}{\gamma_1^2 s^2}
\left(1-\frac{4}{s}\frac{\gamma_2}{\gamma_1}\right).
\label{eq:fuppersb}
\end{alignat}
It remains to address the term of the inelastic peak,
\begin{alignat}{1}
F^{(2)}_{\rm inel}&=
-\eta_2^2\cos^2\phi_2\frac{\Omega_2^2}{2\gamma_2}\langle g\ket{-}^2\times\nonumber\\
&{\rm Re}\sum_{\sigma=\pm}\langle g\ket{\sigma}\obra{\sigma}\rho_{\rm st}
\ket g\frac{(2\langle n\rangle+1)\lambda_{1\sigma}-i\nu}{\lambda_{1\sigma}^2+\nu^2}
\label{eq:finel2}
\end{alignat}
from Eq.~\eqref{eq:peak2}. To arrive at Eq.~\eqref{eq:finel2}, we have taken
into account only the dominating term
$\Tr\{D_0^\dagger\cP_0 D_0\rho_2^{\rm st}\}\propto\gamma_2^{-1}$.
Analogously to the previous treatments, an expansion in $s$ yields
\begin{equation}
F^{(2)}_{\rm inel} =
\eta_2^2\cos^2\phi_2\frac{2\Omega_2^2}{\gamma_1\gamma_2}\frac{\langle n\rangle}{s}
\left(1-\frac{2}{s}\frac{\gamma_2}{\gamma_1}-i\frac{s}{2}\right).
\label{eq:finel2approx}
\end{equation}
When using the approximations for the dark and bright period $T_D$ and $T_B$ from Sec.~\ref{sec:qj}, we eventually find the expressions \eqref{eq:s1inel} and \eqref{eq:s2inel}, in the case $\gamma_2\ll s\gamma_1/2$.


\begin{thebibliography}{99}
\bibitem{Bohr}
N. Bohr, Philos. Mag. {\bf 26}, 1 (1913); 
N. Bohr, Philos. Mag. {\bf 26}, 476 (1913).
\bibitem{qjumpsexp1}
W. Nagourney, J. Sandberg, and H. G. Dehmelt, Phys. Rev. Lett. {\bf 56}, 2797 (1986).
\bibitem{qjumpsexp2}
T. Sauter, W. Neuhauser, R. Blatt, and P. E. Toschek, Phys. Rev. Lett. {\bf 57}, 1696 (1986).
\bibitem{qjumpsexp3}
J. C. Bergquist, R. B. Hulet, W. M. Itano, and D. J. Wineland, Phys. Rev. Lett. {\bf 57}, 1699 (1986).
\bibitem{Leibfried}
D. Leibfried, R. Blatt, C. Monroe and D. Wineland, Rev. Mod. Phys. {\bf 75}, 281 (2003).
\bibitem{Plenio98}
M. B.\ Plenio and P. L.\ Knight, Rev.\ Mod.\ Phys.\ {\bf 70}, 101 (1998).
\bibitem{CarBooks}
H. J. Carmichael, {\it Statistical Methods in Quantum Optics I} and {\it II} (Springer, Berlin 1999 and 2008). 
\bibitem{Cohen86} C. Cohen-Tannoudji and  J. Dalibard, Europhys. Lett. {\bf 1}, 441 (1986).
\bibitem{Carmichael91}
H. Carmichael, {\it An Open Systems Approach to Quantum Optics}, Lecture Notes in Physics (Springer, Berlin 1991).
\bibitem{Molmer}
K. M{\o}lmer, Y. Castin, and J. Dalibard, J. Opt. Soc. Am. B {\bf 10}, 524 (1993).
\bibitem{StochasticCooling}
J. Eschner, B. Appasamy, and P. E. Toschek, Phys. Rev. Lett. {\bf 74}, 2435 (1995).
\bibitem{Almut}
J. Metz, M. Trupke, and A. Beige, Phys. Rev. Lett. 97, 040503 (2006).
\bibitem{Cirac97}
S. A. Gardiner, J. I. Cirac, and P. Zoller, Phys. Rev. A {\bf 55}, 1683 (1997). 
\bibitem{Vogel}
S. Wallentowitz, R. L. de Matos Filho, and W. Vogel, Phys. Rev. A {\bf  56}, 1205 (1997).
\bibitem{Schuck10}
C. Schuck, F. Rohde, N. Piro, M. Almendros, J. Huwer, M. W. Mitchell, M. Heinrich, A. Haase, F. Dubin and J. Eschner, Phys. Rev. A {\bf 81}, 011802 (2010).
\bibitem{Piro10}
N. Piro, F. Rohde, C. Schuck, M. Almendros, J. Huwer, J. Ghosh, A. Haase, M. Hennrich, F. Dubin, and J. Eschner, preprint arXiv:1004.4158 (2010).
\bibitem{Dotsenko09} I. Dotsenko, M. Mirrahimi, M. Brune, S. Haroche, J.-M. Raimond, and P. Rouchon, Phys. Rev. A {\bf 80}, 013805 (2009).
\bibitem{StocCool:exp}
B. Appasamy, Y. Stalgies, and P. E. Toschek, Phys. Rev. Lett. {\bf 80}, 2805 (1998)
\bibitem{Bienert06} M. Bienert, W. Merkel, and G. Morigi,
Phys. Rev. A {\bf 73}, 033402 (2006).
\bibitem{BienertTorres07} M. Bienert, J. M. Torres, S. Zippilli, and G. Morigi,
Phys. Rev. A {\bf 76}, 013410 (2007).
\bibitem{Lindberg86} M. Lindberg,
Phys. Rev. A {\bf 34}, 3178 (1986).
\bibitem{Cirac93} J. I.\ Cirac, R.\ Blatt, A. S.\ Parkins, P.\ Zoller,
Phys.\ Rev.\ A {\bf 48}, 2169  (1993).
\bibitem{Bienert04} M. Bienert, W. Merkel, and G. Morigi,
Phys. Rev. A {\bf 69}, 013405 (2004); see also M. Bienert, PhD-Thesis, University of Ulm, 2004
(Mensch \& Buch Verlag, Berlin 2004, ISBN 3-89820-791-9).
\bibitem{Raab00} Ch.\ Raab, J.\ Eschner, J.\ Bolle, H.\ Oberst,
F.\ Schmidt-Kaler and R.\ Blatt,
Phys.\ Rev.\ Lett.\ {\bf 85}, 538 (2000).
\bibitem{Phillips97} M. Gatzke, G. Birkl, P. S. Jessen, A. Kastberg, S. L. Rolston, and W. D. Phillips, Phys. Rev. A {\bf 55}, R3987 (1997).
\bibitem{Plenio95}
G. C.\ Hegerfeldt and M. B.\ Plenio, Phys.\ Rev.\ A {\bf 52}, 3333 (1995).
\bibitem{Tamm2000}
V. B{\"u}hner and Chr. Tamm, Phys. Rev. A {\bf 61}, 061801 (2000).
\bibitem{Manka1991} A. S. Manka, H. M. Doss, L. M. Narducci, P. Ru and G.-L. Oppo, Phys. Rev. A {\bf 43}, 3748 (1991).
\bibitem{Narducci1990} L. M. Narducci, M. O. Scully, G.-L. Oppo, P. Ru and J. R. Tredicce, Phys. Rev. A {\bf 42}, 1630 (1990).
\bibitem{Stalgies1996} Y. Stalgies, I. Siemers, B. Appasamy, T. Altevogt and P. E. Toschek,
Europhys. Lett. {\bf 35}, 259 (1996).
\bibitem{Gauthier1991}
D. J. Gauthier, Y. Zhu, and T. W. Mossberg,
Phys. Rev. Lett. {\bf 66}, 2460 (1991).
\bibitem{Arimondo:1996}
E. Arimondo, Prog. Optics {\bf 35}, 257 (1996).
\bibitem{Harris:1997}
S. Harris, Phys. Today {\bf 50}, 36 (1997).
\bibitem{Fleischhauer:2005}
M. Fleischhauer, A. Imamoglu, and J. P. Marangos, Rev. Mod. Phys., {\bf 77}, 633 (2005).
\bibitem{Aspect1984} A. Aspect, J. Dalibard, P. Grangier and G. Roger,
Opt. Commun., {\bf 49}, 429 (1984).
\bibitem{Zhou:1997}
P. Zhou and S. Swain, Phys. Rev. Lett. {\bf 77}, 3995 (1996).
\bibitem{Martinez:1997}
M. A. G. Martinez, P. R. Herczfeld, C. Samuels, L. M. Narducci and C. H. Keitel, Phys. Rev. A, {\bf 55}, 4483 (1997).
\bibitem{cohenbook} C. Cohen-Tannoudji, J. Dupont-Roc, and G. Grynberg, {\it Atom-Photon-Interactions} (Wiley-VCH, Weinheim 2004).
\bibitem{Itano90}
W. M. Itano, D. J. Heinzen, J. J. Bollinger, and D. J. Wineland, Phys. Rev. A {\bf 41}, 2295 (1990).
\bibitem{Eschner03} J. Eschner and G. Morigi and F. Schmidt-Kaler and R. Blatt,
J. Opt. Soc. Am. B {\bf 20}, 1003 (2003).
\bibitem{Stenholm86} S. Stenholm, Rev. Mod. Physics, {\bf 58}, 699 (1986).
\bibitem{Cirac92} J. I. Cirac, R. Blatt, P. Zoller, and W. D. Phillips, Phys. Rev. A {\bf 46}, 2668 (1992).
\bibitem{Englert2002} 
B.-G.\ Englert and G.\ Morigi, in {\it Coherent Evolution in Noisy
Environments}, Lecture Notes in Physics {\bf 611}, p.\ 55, ed.\ by
A. Buchleitner, K. Hornberger (Springer Verlag,
Berlin-Heidelberg-New York 2002).
\bibitem{Mollow69} B. R. Mollow, Phys. Rev.  {\bf 188}, 1969 (1969).
\bibitem{Mundt}
A. Kreuter, C. Becher, G. P. T. Lancaster, A. B. Mundt, C. Russo, H. H\"affner, C. Roos, J. Eschner, F. Schmidt-Kaler, and R. Blatt, Phys. Rev. Lett. {\bf 92}, 203002 (2004).
\bibitem{Jakob03} M. Jakob and S. Stenholm, Phys. Rev. A {\bf 67}, 032111 (2003).
\bibitem{Metz2007a}J. Metz, C. Sch\"on, and A. Beige
Phys. Rev. A {\bf 76}, 052307 (2007).
\bibitem{Metz2007b}J. Metz and A. Beige
Phys. Rev. A {\bf 76}, 022331 (2007).
\end{thebibliography}
\end{document}